\def\etc{{\it etc.}}
\def\etal{{\it et al.}}

\def\~{{$\tilde{\phantom{a}}$}}

\documentclass [12pt] {article}
\usepackage{epsfig}
\usepackage{color}

\def\thebibliography#1{\section{References}\markboth
 {REFERENCES}{REFERENCES}\list
 {[\arabic{enumi}]}{\settowidth\labelwidth{[#1]}\leftmargin\labelwidth
 \advance\leftmargin\labelsep
 \usecounter{enumi}}
 \def\newblock{\hskip .11em plus .33em minus -.07em}
 \sloppy
 \sfcode`\.=1000\relax}
\def\upcite#1{\raise6pt\hbox{\scriptsize
\cite{#1}}}
  \def\lsim{\mathrel {\vcenter {\baselineskip 0pt \kern 0pt
    \hbox{$<$} \kern 0pt \hbox{$\sim$} }}}
    \def\gsim{\mathrel {\vcenter {\baselineskip 0pt \kern 0pt
    \hbox{$>$} \kern 0pt \hbox{$\sim$} }}}

\def\hline{\noalign{\hrule \vskip2pt}}

\def\|{\ifmmode\Vert\else \char`\|\fi}
\ifx\oldzeta\undefined                          
  \let\oldzeta=\zeta                            
  \def\zzeta{{\raise 2pt\hbox{$\oldzeta$}}}     
  \let\zeta=\zzeta                              
\fi

\ifx\oldchi\undefined                           
  \let\oldchi=\chi                              
  \def\cchi{{\raise 2pt\hbox{$\oldchi$}}}       
  \let\chi=\cchi                                
\fi

\def\frac#1#2{{#1 \over #2}}

\def\half{\ifinner {\scriptstyle {1 \over 2}}
   \else {1 \over 2} \fi}

\def\abs#1{\left\vert#1\right\vert}	

\def\simge{\mathrel{%
   \rlap{\raise 0.511ex \hbox{$>$}}{\lower 0.511ex \hbox{$\sim$}}}}
\def\simle{\mathrel{
   \rlap{\raise 0.511ex \hbox{$<$}}{\lower 0.511ex \hbox{$\sim$}}}}

\def\buildchar#1#2#3{{\null\!                   
   \mathop#1\limits^{#2}_{#3}                   
   \!\null}}                                    
\def\overcirc#1{\buildchar{#1}{\circ}{}}

\def\slashchar#1{\setbox0=\hbox{$#1$}           
   \dimen0=\wd0                                 
   \setbox1=\hbox{/} \dimen1=\wd1               
   \ifdim\dimen0>\dimen1                        
      \rlap{\hbox to \dimen0{\hfil/\hfil}}      
      #1                                        
   \else                                        
      \rlap{\hbox to \dimen1{\hfil$#1$\hfil}}   
      /                                         
   \fi}                                         %

\def\subrightarrow#1{
  \setbox0=\hbox{
    $\displaystyle\mathop{}
    \limits_{#1}$}
  \dimen0=\wd0
  \advance \dimen0 by .5em
  \mathrel{
    \mathop{\hbox to \dimen0{\rightarrowfill}}
       \limits_{#1}}}                           

\def\overlay#1#2{\ifmmode%
\setbox0=\hbox{$#1$}%
\setbox1=\hbox to\wd0{\hss$#2$\hss}\else%
\setbox0=\hbox{#1}%
\setbox1=\hbox to\wd0{\hss#2\hss}\fi%
#1\hskip-\wd0\box1 }

\def\pmb#1{\leavevmode\setbox0=\hbox{#1}%
\kern-.02em\copy0\kern-\wd0
\kern.04em\copy0\kern-\wd0
\kern-.02em\raise.04em\box0 }

\def\vereq#1#2{\lower3pt\vbox{\baselineskip1.5pt \lineskip1.5pt
\ialign{$\m@th#1\hfill##\hfil$\crcr#2\crcr\sim\crcr}}}

\def\tensor#1{\protect\@ontopof{#1}{\leftrightarrow}{1.15}\mathord{\box2}}
\def\overstar#1{\protect\@ontopof{#1}{\ast}{1.15}\mathord{\box2}}
\def\overdots#1{\protect\@ontopof{#1}{\cdots}{1.0}\mathord{\box2}}
\def\overcirc#1{\protect\@ontopof{#1}{\circ}{1.2}\mathord{\box2}}
\def\loarrow#1{\protect\@ontopof{#1}{\leftarrow}{1.15}\mathord{\box2}}
\def\roarrow#1{\protect\@ontopof{#1}{\rightarrow}{1.15}\mathord{\box2}}

\def\@ontopof#1#2#3{%
{\mathchoice
{\@@ontopof{#1}{#2}{#3}\displaystyle\scriptstyle}%
{\@@ontopof{#1}{#2}{#3}\textstyle\scriptstyle}%
{\@@ontopof{#1}{#2}{#3}\scriptstyle\scriptscriptstyle}%
{\@@ontopof{#1}{#2}{#3}\scriptscriptstyle\scriptscriptstyle}%
}%
}

\def\@@ontopof#1#2#3#4#5{%
\setbox0=\hbox{$#4#1$}%
\setbox1=\hbox{$#5#2$}%
\setbox2=\hbox{}\ht2=\ht0 \dp2=\dp0 %
\ifdim\wd0>\wd1 %
\setbox1=\hbox to\wd0{\hss\box1\hss}%
\mathord{\rlap{\raise#3\ht0\box1}\box0}%
\else   %
\setbox1=\hbox to.9\wd1{\hss\box1\hss}%
\setbox0=\hbox to\wd1{\hss$#4\relax#1$\hss}%
\mathord{\rlap{\copy0}\raise#3\ht0\box1}%
\fi
}%

\def\lambdabar{\protect\@lambdabar}
\def\@lambdabar{%
\relax
\bgroup
\def\@tempa{\hbox{\raise.73\ht0
\hbox to0pt{\kern.25\wd0\vrule width.5\wd0
height.1pt depth.1pt\hss}\box0}}%
\mathchoice{\setbox0\hbox{$\displaystyle\lambda$}\@tempa}%
{\setbox0\hbox{$\textstyle\lambda$}\@tempa}%
{\setbox0\hbox{$\scriptstyle\lambda$}\@tempa}%
{\setbox0\hbox{$\scriptscriptstyle\lambda$}\@tempa}%
\egroup
}

\def\corresponds{{\lower.2ex\hbox{=}}{\rm\kern-.75em^\triangle}}
\def\succsim{\succ\kern-.9em_\sim\kern.3em}
\def\precsim{\prec\kern-1em_\sim\kern.3em}
\def\slantfrac#1#2{\kern1em^{#1}\kern-.3em/\kern-.1em_{#2}}

\begin{document}
                                                                
\begin{center}
{\Large\bf Bessel Beams}
\\

\medskip

Kirk T.~McDonald
\\
{\sl Joseph Henry Laboratories, Princeton University, Princeton, NJ 08544}
\\
(June 17, 2000)
\end{center}

\section{Problem}

Deduce the form of a cylindrically symmetric plane electromagnetic
wave that propagates in vacuum.  

A scalar, azimuthally symmetric wave of frequency $\omega$
that propagates in the positive $z$ direction could be written as
\begin{equation}
\psi({\bf r},t) = f(\rho) e^{i(k_z z - \omega t)},
\label{eq1}
\end{equation}
where $\rho = \sqrt{x^2 + y^2}$.  Then, the problem is to deduce the
form of the radial function $f(\rho)$ and any relevant condition on
the wave number $k_z$, and to relate that scalar wave function to
a complete solution of Maxwell's equations.

The waveform (\ref{eq1}) has both wave velocity and group velocity equal to
$\omega / k_z$.  Comment on the apparent superluminal character of
the wave in case that $k_z < k = \omega / c$, where $c$ is the speed
of light.

\section{Solution}

As the desired solution for the radial wave function proves to be a
Bessel function, the cylindrical plane waves have come to be called
Bessel beams, following their introduction by Durnin \etal \
\cite{Durnin1,Durnin2}.  The question of superluminal behavior of
Bessel beams has recently been raised by Mugnai \etal \
\cite{Mugnai}.

Bessel beams are a realization of super-gain antennas
\cite{Schelkunoff,Bouwkamp,Yaru} in the optical domain.
A simple experiment to generate Bessel beams is described in \cite{McQueen}.

Sections 2.1 and 2.2 present two methods of solution for Bessel beams
that satisfy the Helmholtz wave equation.  The issue of group and signal
velocity for these waves is discussed in sec.~2.3.
Forms of Bessel beams that satisfy Maxwell's equations are given in 
sec.~2.4.

\subsection{Solution via the Wave Equation}

On substituting the form (\ref{eq1}) into the wave equation,
\begin{equation}
\nabla^2 \psi = { 1 \over c^2} {\partial^2 \psi \over \partial t^2},
\label{eq2}
\end{equation}
we obtain
\begin{equation}
{d^2 f \over d\rho^2} + {1 \over \rho} {d f \over d \rho} +
(k^2 - k_z^2) f = 0.
\label{eq3}
\end{equation}
This is the differential equation for Bessel functions of order 0,
so that
\begin{equation}
f(\rho) = J_0(k_r \rho),
\label{eq4}
\end{equation}
where
\begin{equation}
k_\rho^2 + k_z^2 = k^2.
\label{eq5}
\end{equation}

The form of eq.~(\ref{eq5}) suggests that we introduce a (real)
parameter $\alpha$ such that
\begin{equation}
k_\rho = k \sin \alpha, \qquad \mbox{and} \qquad k_z = k \cos\alpha.
\label{eq6}
\end{equation}
Then, the desired cylindrical plane wave has the form
\begin{equation}
\psi({\bf r},t) = J_0(k \sin\alpha \, \rho) 
e^{i(k \cos\alpha \, z - \omega t)},
\label{eq7}
\end{equation}
which is commonly called a Bessel beam.
The physical significance of parameter $\alpha$, and that of the
group velocity 
\begin{equation}
v_g = {d \omega \over d k_z} = {\omega \over k_z} = v_p = {c \over \cos\alpha}
\label{eq8}
\end{equation}
will be discussed in sec.~2.3.

While eq.~(\ref{eq7}) is a solution of the Helmholtz wave equation
(\ref{eq2}), assigning $\psi({\bf r},t)$ to be a single component of an
electric field, say $E_x$, does not provide a full solution to Maxwell's
equations.  For example, if ${\bf E} = \psi \hat{\bf x}$, then
$\nabla \cdot {\bf E} = \partial \psi / \partial x \neq 0$.  
Bessel beams that satisfy Maxwell's equations are given in sec.~2.4.

\subsection{Solution via Scalar Diffraction Theory}

The Bessel beam (\ref{eq7}) has large amplitude only for $\abs{\rho}
\lsim 1/ k \sin\alpha$, and maintains the same radial profile over
arbitrarily large propagation distance $z$.  This behavior appears to
contradict the usual lore that a beam of minimum transverse extent 
$a$ diffracts to fill a cone of angle $1/a$.  Therefore, the Bessel
beam (\ref{eq7}) has been called ``diffraction free'' \cite{Durnin2}.

Here, we show that the Bessel beam does obey the formal laws of
diffraction, and can be deduced from scalar diffraction theory.

According to that theory \cite{Jackson}, a cylindrically symmetric
wave $f(\rho)$ of frequency $\omega$ at the plane $z = 0$ propagates
to point {\bf r} with amplitude
\begin{equation}
\psi({\bf r},t) = {k \over 2 \pi i} \int \int \rho' d\rho' d\phi f(\rho')
{e^{i(k R - \omega t)} \over R},
\label{eq9}
\end{equation}
where $R$ is the distance between the source and observation point.
Defining the observation point to be $(\rho,0,z)$, we have
\begin{equation}
R^2 =z^2 + \rho^2 + \rho^{'2} - 2 \rho \rho' \cos\phi,
\label{eq10}
\end{equation}
so that for large $z$,
\begin{equation}
R \approx z + {\rho^2 + \rho^{'2} - 2 \rho \rho' \cos\phi \over 2 z}.
\label{eq11}
\end{equation}

In the present case, we desire the amplitude to have form (\ref{eq1}).
As usual, we approximate $R$ by $z$ in the denominator of 
eq.~(\ref{eq9}), while using approximation (\ref{eq11}) in the 
exponential factor.  This leads to the integral equation
\begin{eqnarray}
f(\rho) e^{i k_z z} & = & {k \over 2 \pi i} {e^{ik z}
e^{i k \rho^2 / 2 z}
\over z} \int_0^\infty \rho' d\rho' f(\rho') e^{i k \rho^{'2} / 2z}
\int_0^{2 \pi} d\phi e^{-i k \rho \rho' \cos\phi / z}
\nonumber \\
& = & {k \over i} {e^{ik z}
e^{i k \rho^2 / 2 z}
\over z} \int_0^\infty \rho' d\rho' f(\rho') J_0(k \rho \rho' / z)
e^{i k \rho^{'2} / 2z},
\label{eq12}
\end{eqnarray}
using a well-known integral representation of the Bessel function
$J_0$.

It is now plausible that the desired eigenfunction $f(\rho)$ is a
Bessel function, say $J_0(k_\rho \rho)$,
and on consulting a table of integrals of Bessel
functions we find an appropriate relation \cite{Gradshteyn},
\begin{equation}
\int_0^\infty \rho' d\rho' J_0(k_{\rho} \rho') J_0(k \rho \rho' / z)
e^{i k \rho^{'2} / 2z} = {i z \over k} e^{-i k \rho^2 / 2 z}
e^{- i k_\rho^2 z / 2 k} J_0(k_\rho \rho).
\label{eq13}
\end{equation}
Comparing this with eq.~(\ref{eq12}), we see that $f(\rho) = 
J_0(k_\rho \rho)$ is indeed an eigenfunction provided that
\begin{equation}
k_z = k - {k_\rho^2 \over 2 k}.
\label{eq14}
\end{equation}
Thus, if we write $k_\rho = k \sin\alpha$, then for small $\alpha$,
\begin{equation}
k_z \approx k (1 - \alpha^2 / 2) \approx k \cos\alpha,
\label{eq15}
\end{equation}
and the desired cylindrical wave again has form (\ref{eq7}).

Strictly speaking, the scalar diffraction theory reproduces the
``exact'' result (\ref{eq7}) only for small $\alpha$.  But the scalar
diffraction theory is only an approximation, and we predict with
confidence that an ``exact'' diffraction theory would lead to the
form (\ref{eq7}) for all values of parameter $\alpha$.  That is,
``diffraction-free'' beams are predicted within diffraction theory.

It remains that the theory of diffraction predicts that an infinite aperture is
needed to
produce a beam whose transverse profile is invariant with longitudinal
distance.  That a Bessel beam is no exception to this rule is reviewed in
sec.~2.3.

The results of this section were inspired by \cite{Jiang}.  One of the
first solutions for Gaussian laser beams was based on scalar diffraction
theory cast as an eigenfunction problem \cite{Boyd}.

\subsection{Superluminal Behavior}

In general, the group velocity (\ref{eq8}) of a Bessel beam exceeds
the speed of light.  However, this apparently superluminal behavior
cannot be used to transmit signals faster than lightspeed.

An important step towards understanding this comes from the
interpretation of parameter $\alpha$ as the angle with respect to
the $z$ axis of the wave vectors of an infinite set of ordinary
plane waves whose superposition yields the Bessel beam \cite{Eberly}.
To see this, we invoke the integral representation of the Bessel
function  to write eq.~(\ref{eq7}) as
\begin{eqnarray}
\psi({\bf r},t) & =  &J_0(k \sin\alpha \, \rho) 
e^{i(k \cos\alpha \, z - \omega t)} 
\nonumber \\
& = & {1 \over 2 \pi} \int_0^{2 \pi} d \phi 
e^{i(k \sin\alpha \, x \cos\phi + k \sin\alpha \, y \sin\phi
+ k \cos\alpha \, z - \omega t)}
\label{eq16} \\
& = & {1 \over 2 \pi} \int_0^{2 \pi} d \phi 
e^{i({\bf q} \cdot {\bf r} - \omega t)},
\nonumber
\end{eqnarray}
where the wave vector {\bf q}, given by
\begin{equation}
{\bf q} = k (\sin\alpha \cos\phi, \sin\alpha \sin\phi, \cos\alpha),
\label{eq17}
\end{equation}
makes angle $\alpha$ to the $z$ axis as claimed.

We now see that a Bessel beam is rather simple to produce in 
principle \cite{Durnin2}.  Just superpose all possible plane waves
with equal amplitude and a common phase 
that make angle $\alpha$ to the $z$ axis, 

According to this prescription, we expect the $z$ axis to be uniformly
illuminated by the Bessel beam.  If that beam is created at the plane
$z = 0$, then any annulus of equal radial extent in that plane must
project equal power into the beam.  For large $\rho$ this is readily
confirmed by noting that $J_0^2(k \sin\alpha\, \rho) \approx
\cos^2(k \sin\alpha\, \rho + \delta)/ (k \sin\alpha\, \rho)$, so the
integral of the power over an annulus of one radial period, $\Delta \rho =
\pi / (k \sin\alpha)$, is independent of radius.

Thus, from an energy perspective a Bessel beam is not confined to a finite
region about the $z$ axis.  If the beam is to propagate a distance $z$
from the plane $z = 0$, it must have radial extent of at least $ \rho =
z \tan\alpha$ at $z = 0$.  An arbitrarily large initial aperture, and
arbitrarily large power, is
required to generate a Bessel beam that retains its ``diffraction-free''
character over an arbitrarily large distance.

Each of the plane waves that makes up the Bessel beam propagates with
velocity $c$ along a ray that makes
angle $\alpha$ to the $z$ axis.  The intersection of the $z$ axis and
a plane of constant phase of any of these wave moves forward with
superluminal speed $c / \cos\alpha$, which is equal to the phase and group 
velocities (\ref{eq8}).

This superluminal behavior does not represent any violation of special 
relativity, but is an example of the ``scissors paradox" that the point of
contact of a pair of scissors could move faster than the speed of light
while the tips of the blades are moving together at sublightspeed.
A ray of sunlight that makes angle $\alpha$ to the surface of the
Earth similarly leads to a superluminal velocity $c / \cos\alpha$
of the point of contact of a wave front with the Earth.

However, we immediately see that a Bessel beam could not be used to
send a signal from, say, the origin, $(0,0,0)$, to a point $(0,0,z)$ at
 a speed faster than light.
A Bessel beam at $(0,0,z)$ is made of rays of plane waves that
intersect the plane $z = 0$
at radius $\rho = z \tan\alpha$.  Hence, to deliver a message from
$(0,0,0)$ to $(0,0,z)$ via a
Bessel beam, the information must first propagate from the origin
out to at least radius $\rho = z \tan\alpha$ at $z = 0$ to set up 
the beam.
Then, the rays must propagate distance $z/\cos\alpha$ to reach point
$z$ with the message.  The total distance traveled by the
information is thus $z(1 + \sin\alpha)/\cos\alpha$, and the signal
velocity $v_s$ is given by
\begin{equation}
v_s \approx c {\cos\alpha \over 1 + \sin\alpha},
\label{eq18}
\end{equation}
 which is
always less than $c$.  The group velocity and signal velocity for a
Bessel beam are very different.  
Rather than being a superluminal carrier of information at its group
velocity $c / \cos\alpha$, a modulated Bessel beam could be used to
deliver messages only at speeds well below that of light.

\subsection{Solution via the Vector Potential}

To deduce all components of the electric and magnetic fields of
a Bessel beam that satisfies Maxwell's equation starting from a single scalar 
wave function, we follow the 
suggestion of Davis \cite{Davis} and seek solutions for a vector potential 
{\bf A} that has only a single component.  We work in the Lorentz gauge
(and Gaussian units), so that the scalar potential $\Phi$ is related by
\begin{equation}
\nabla \cdot {\bf A} + {1 \over c} {\partial \Phi \over \partial t} = 0.
\label{e1}
\end{equation}
The vector potential can therefore have a nonzero divergence, which permits
solutions having only a single component.  Of course, the electric and
magnetic fields can be deduced from the potentials via
\begin{equation}
{\bf E} = - \nabla \Phi - {1 \over c} {\partial {\bf A} \over \partial t},
\label{e2}
\end{equation} 
and
\begin{equation}
{\bf B} = \nabla \times {\bf A}.
\label{e3}
\end{equation}
For this, the scalar potential must first be deduced from the vector 
potential using the Lorentz condition (\ref{e1}).  We consider waves of
frequency $\omega$ and time dependence of the form $e^{-i \omega t}$, so
that $\partial \Phi / \partial t = - i k \Phi$.  Then, the Lorentz
condition yields
\begin{equation}
\Phi = - {i \over k} \nabla \cdot {\bf A},
\label{e4}
\end{equation}
and the electric field is given by
\begin{equation}
{\bf E} = ik \left[ {\bf A} + {1 \over k^2} {\bf \nabla} ({\bf \nabla}
\cdot {\bf A}) \right].
\label{e5}
\end{equation}
Then, $\nabla \cdot {\bf E} = 0$ since $\nabla^2 (\nabla \cdot {\bf A}) +
k^2 (\nabla \cdot {\bf A}) = 0$ for a vector potential {\bf A} of frequency
$\omega$ that satifies the wave equation (\ref{eq2}), \etc

We already have a scalar solution (\ref{eq7}) to the wave equation, which
we now interpret as the only nonzero component, $A_j$, of the vector
potential for a Bessel beam that propagates in the $+z$ direction,
\begin{equation}
A_j({\bf r},t) = \psi({\bf r},t) 
 \propto J_0(k \sin\alpha\, \rho) e^{i(k \cos\alpha\, z - \omega t)}.
\label{e6}
\end{equation}

We consider five choices for the meaning of index $j$, namely $x$, $y$,
$z$, $\rho$, and $\phi$, which lead to five types of Bessel beams.  Of
these, only the case of $j = z$ corresponds to physical, azimuthally
symmetric fields, and so perhaps should be called the Bessel beam.

\subsubsection{$j = x$}

In this case,
\begin{equation}
\nabla \cdot {\bf A} = {\partial \psi \over \partial x} = 
- {k \sin\alpha \, x \over \rho} J_1(k \sin\alpha\, \rho)
e^{i(k \cos\alpha\, z - \omega t)}.
\label{e7}
\end{equation}
In calculating $\nabla(\nabla \cdot {\bf A})$ we use the identity
$J_1' = (J_0 - J_2)/2$.  Also, we divide {\bf E} and {\bf B} by the 
factor $ik$ to present the results in a simpler form.  We find,
\begin{eqnarray}
E_x & = & \left\{ J_0(\varrho) -  {\sin^2\alpha\ \over \rho^2}
\left[  
 {y^2 J_1(\varrho) \over \varrho}
- {x^2 \over 2} \left(J_0(\varrho) - J_2(\varrho) \right) \right] \right\}
e^{i(k \cos\alpha\, z - \omega t)},
\nonumber \\
E_y & = & {\sin^2\alpha\, x y \over \rho^2} \left[ 
{ J_1(\varrho) \over \varrho}
- {1 \over 2} \left(J_0(\varrho) - J_2(\varrho) \right) \right]
e^{i(k \cos\alpha\, z - \omega t)},
\label{e8} \\
E_z & = & - i \sin 2\alpha {x \over 2 \rho} J_1(\varrho)
e^{i(k \cos\alpha\, z - \omega t)},
\nonumber
\end{eqnarray}
where
\begin{equation}
\varrho \equiv k \sin\alpha\, \rho,
\label{e9}
\end{equation}
and
\begin{eqnarray}
B_x & = & 0,
\nonumber \\
B_y & = & \cos\alpha\, J_0(\varrho) e^{i(k \cos\alpha\, z - \omega t)},
\label{e10} \\
B_z & = & - i \sin\alpha {x \over  \rho} J_1(\varrho)
e^{i(k \cos\alpha\, z - \omega t)}.
\nonumber
\end{eqnarray}

A Bessel beam that obeys Maxwell's equations and has purely $x$
polarization of its electric field on the
$z$ axis includes nonzero $y$ and $z$ polarization at points off that
axis, and does not exhibit the azimuthal symmetry of the underlying
vector potential.

\subsubsection{$j = y$}

This case is very similar to that of $j = x$.

\subsubsection{$j = z$}

In this case the electric and magnet fields retain azimuthal symmetry,
so that it is convenient to display the $\rho$, $\phi$ and $z$ components
of the fields.  First,
\begin{equation}
\nabla \cdot {\bf A} = {\partial \psi \over \partial z} = 
i k \cos\alpha \, J_0(k \sin\alpha\, \rho)
e^{i(k \cos\alpha\, z - \omega t)}.
\label{e11}
\end{equation}
Then, we divide the electric and magnetic fields by $k \sin\alpha$ to
find the relatively simple forms:
\begin{eqnarray}
E_\rho & = & \cos\alpha\, J_1(\varrho)
e^{i(k \cos\alpha\, z - \omega t)},
\nonumber \\
E_\phi & = & 0,
\label{e12} \\
E_z & = & i \sin\alpha\,  J_0(\varrho) 
e^{i(k \cos\alpha\, z - \omega t)},
\nonumber
\end{eqnarray}
and 
\begin{eqnarray}
B_\rho & = & 0,
\nonumber \\
B_\phi & = & J_1(\varrho) e^{i(k \cos\alpha\, z - \omega t)},
\label{e13} \\
B_z & = & 0.
\nonumber
\end{eqnarray}
This Bessel beam is a transverse magnetic (TM) wave.  
The radial electric field $E_\rho$ vanishes on the $z$ axis 
(as it must if that axis is
charge free), while the longitudinal electric field $E_z$ is maximal
there.  Cylindrically symmetric waves  with radial electric polarization
are often called axicon beams \cite{McLeod}. 

\subsubsection{$j = \rho$}

In this case,
\begin{equation}
\nabla \cdot {\bf A} = {1 \over \rho} {\partial \rho \psi \over 
\partial \rho} = \left[ {J_0(k \sin\alpha\, \rho) \over \rho} 
- k \sin\alpha\, J_1(k \sin\alpha\, \rho) \right]
e^{i(k \cos\alpha\, z - \omega t)}.
\label{e14}
\end{equation}
After dividing by $ik$, the electric and magnetic fields are
\begin{eqnarray}
E_\rho & = & \left\{ J_0(\varrho) - \sin^2\alpha \left[
{J_0(\varrho) \over \varrho^2} + {J_1(\varrho) \over \varrho}
+ {1 \over 2} (J_0(\varrho - J_2(\varrho)) \right] \right\}
e^{i(k \cos\alpha\, z - \omega t)},
\nonumber \\
E_\phi & = & 0,
\label{e15} \\
E_z & = & i \cos\alpha \sin\alpha \left[  {J_0(\varrho) \over \varrho} 
- J_1(\varrho) \right] e^{i(k \cos\alpha\, z - \omega t)},
\nonumber
\end{eqnarray}
and 
\begin{eqnarray}
B_\rho & = & 0,
\nonumber \\
B_\phi & = & \cos\alpha\, J_0(\varrho) e^{i(k \cos\alpha\, z - \omega t)},
\label{e16} \\
B_z & = & 0.
\nonumber
\end{eqnarray}
The radial electric field diverges as $1 / \rho^2$ for small $\rho$, so
this case is unphysical.

\subsubsection{$j = \phi$}

Here,
\begin{equation}
\nabla \cdot {\bf A} = {1 \over \rho} {\partial \psi \over 
\partial \phi} = 0.
\label{e17}
\end{equation}
After dividing by $ik$, the electric and magnetic fields are
\begin{eqnarray}
E_\rho & = & 0,
\nonumber \\
E_\phi & = & J_0(\varrho) e^{i(k \cos\alpha\, z - \omega t)},
\label{e18} \\
E_z & = & 0,
\nonumber
\end{eqnarray}
and 
\begin{eqnarray}
B_\rho & = & - \cos\alpha\, J_0(\varrho) e^{i(k \cos\alpha\, z - \omega t)},
\nonumber \\
B_\phi & = & 0,
\label{e19} \\
B_z & = & - i \sin\alpha\ \left[ {J_0(\varrho) \over \varrho}
- J_1(\varrho) \right] e^{i(k \cos\alpha\, z - \omega t)}.
\nonumber
\end{eqnarray}
These fields are unphysical due to the finite value of $E_\phi$ at
$\rho = 0$, and the divergence of $B_z$ as $\rho \to 0$.

\end{document}